%% file: acl_latex.tex
\pdfoutput=1

\documentclass[11pt]{article}

\usepackage[preprint]{acl}

\usepackage{times}
\usepackage{latexsym}

\usepackage[T1]{fontenc}

\usepackage{listings}
\usepackage[utf8]{inputenc}

\usepackage{microtype}

\usepackage{inconsolata}

\usepackage{graphicx}

\usepackage{amsmath}
\usepackage{amssymb}
\usepackage{adjustbox}
\usepackage{microtype}
\usepackage{float}
\usepackage{hyperref}
\usepackage{url}
\usepackage{booktabs}
\usepackage{geometry}
\usepackage{multirow}
\usepackage{algorithm}
\usepackage{algorithmic}
\usepackage{wrapfig}
\usepackage{subcaption}
\usepackage{amsmath}
\usepackage{listings}
\usepackage{natbib}
\usepackage{xspace}
\NewDocumentCommand{\heng}
{ mO{} }{\textcolor{red}{\textsuperscript{\textit{Heng}}\textsf{\textbf{\small[#1]}}}}

\NewDocumentCommand{\xiaomeng}
{ mO{} }{\textcolor{green}{\textsuperscript{\textit{Xiaomeng}}\textsf{\textbf{\small[#1]}}}}

\input{def}

\title{REAL: Realism Evaluation of Text-to-Image Generation Models for Effective Data Augmentation}
\author{Ran Li$^1$, Xiaomeng Jin$^2$, Heng Ji$^2$,\\
Columbia University$^1$, University of Illinois Urbana-Champaign$^2$ \\
\texttt{rl3424@columbia.edu, \{xjin17, hengji\}@illinois.edu}
}

\begin{document}
\maketitle

\input{latex/abstract}
\input{latex/introduction}
\input{latex/related_work}
\input{latex/method}
\input{latex/experiments}
\input{latex/discussion}
\input{latex/conclusion}
\input{latex/limitations}

\bibliography{custom}

\appendix
\input{latex/appendix}

\end{document}

%% file: def.tex
\def\name{\textit{REAL}\xspace}

\def\0{{\bf 0}}
\def\1{{\bf 1}}

%% file: latex/abstract.tex
\begin{abstract}
    Recent advancements in text-to-image (T2I) generation models have transformed the field. However, challenges remain in generating images that reflect demanding textual descriptions, especially for fine-grained details and unusual relationships. Existing evaluation metrics focus on text-image alignment but overlook the realism of the generated image, which can be crucial for downstream applications like data augmentation in machine learning. To address this gap, we propose \name, an automatic evaluation framework that assesses realism of T2I outputs along three dimensions: fine-grained visual attributes, unusual visual relationships, and visual styles. \name achieves a Spearman’s $\rho$ score of up to 0.62 in alignment with human judgment and demonstrates utility in ranking and filtering augmented data for tasks like image captioning, classification, and visual relationship detection. Empirical results show that high-scoring images evaluated by our metrics improve F1 scores of image classification by up to 11.3\%, while low-scoring ones degrade that by up to 4.95\%. We benchmark four major T2I models across the realism dimensions, providing insights for future improvements in T2I output realism.
\end{abstract}

%% file: latex/introduction.tex
\section{Introduction}
\begin{figure*}[ht!]
    \centering
    \includegraphics[width=\textwidth]{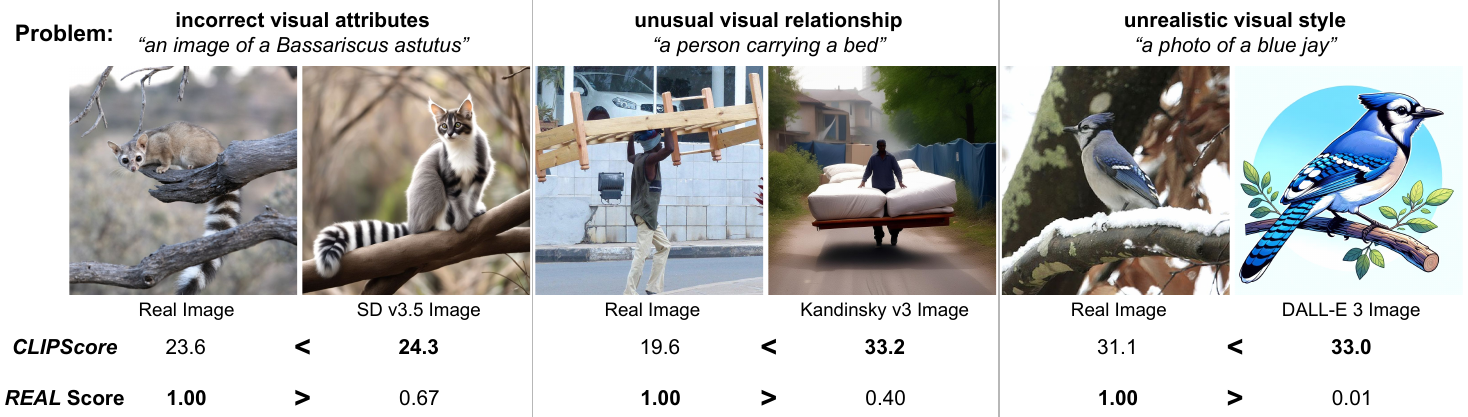}
    \caption{Overview of the three dimensions of realism we study. On the left, a image generated by Stable Diffusion v3.5 exhibits incorrect visual attributes for the species, resembling a cat despite having the distinctive tail. In the middle, Kandinsky 3 struggles with the unusual visual relationships, generating a person that's overlapping with the bed instead of carrying it. On the right, DALL-E 3 produces a stylized illustration instead of a photorealistic image as prompted. \name score correctly identifies all real images, whereas CLIPScore does not.}
    \label{fig:teaser}
\end{figure*}

\begin{figure*}[ht!]
    \centering
    \includegraphics[width=\textwidth]{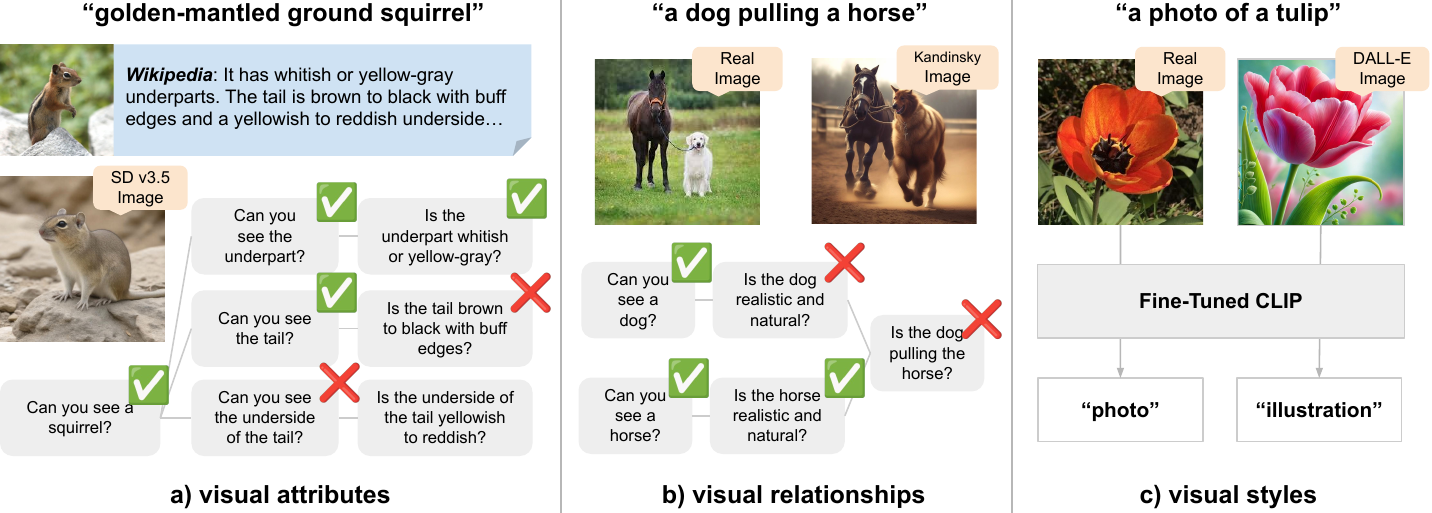}
    \caption{Overview of the three components of the \name framework. For visual attributes and relationships, \name performs schematic evaluation on the presense and realism of each component. For visual styles, \name leverages a fine-tuned CLIP model for rating.}
    \label{fig:method}
\end{figure*}

Recent years have witnessed significant development in T2I generation models \cite{reed2016generative, xu2018attngan, sd1, dalle3, kandinsky3, sd3}. Nevertheless, challenges persist in generating images that accurately reflect demanding textual descriptions.
As shown in Figure~\ref{fig:teaser} and \ref{fig:method}, this is especially true for prompts that involve objects with fine-grained details and intricate relationships between them. Several metrics have been developed to evaluate the faithfulness of generated images to their textual prompts. One of the most popular metric was CLIPScore \cite{hessel2021clipscore}, but the CLIP model it leverages is unreliable for more complex tasks such as visual reasoning \cite{clip}. To overcome this weakness, \citet{hu2023tifa} proposed TIFA, which structurally prompts a visual question answering (VQA) model for more detailed evaluation. Furthermore, Davisonian scene graph \cite{chodavidsonian} defines a detailed standard for schemed evaluation, offering insights in how the evaluation prompts should be generated to ensure reliable results.

While the above-mentioned evaluation frameworks are effective in assessing an output image's faithfulness to its textual prompt, they usually ignore to assess the realism of the generated images. It has been shown that T2I models can be applied to generate augmented data for various machine learning tasks \cite{shivashankar2023semantic, jin2024schema, jin2024armada}. For this purpose, not only faithfulness but also the realism of these images is crucial for training effective downstream machine learning models. High-quality data is essential for model performance, as poor data quality can lead to biased or inaccurate models \cite{polyzotis2019data, jain2020overview}. Despite the importance of image realism, there is a notable lack of automatic evaluation frameworks specifically designed to assess the realism of synthetic images used in data augmentation.

To bridge this gap, we propose \name, a framework for evaluating output realism of T2I models. Distinct from previous works, we measure three unique aspects of realism: correctness of fine-grained visual attributes, plausibility of unusual visual relationships, and realistic visual styles, through prompting a VQA model based on sophisticated schemas.
\name demonstrates strong alignment with human judgment with a Spearman's $\rho$ score of up to 0.62, and we furthermore leverage its output scores to rank and filter augmented data for several downstream computer vision tasks, including image captioning, classification, and visual relationship detection. Empirical results indicate that images assigned high scores by our framework lead to improved training outcomes. For example, augmenting the training set with images labeled with high realism score by \name could enhance F1 score for image classification by up to 11.3\%, and using the ones with low realism scores could deteriorate it by up to 4.95\%. Finally, we benchmark four major T2I models on the three dimensions, providing insights in each model's strength and weakness that guide future researchers to improve output realism of T2I models.

Our contributions are as follows:
\begin{itemize}
    \item Unique from previous works that evaluate T2I alignment, we evaluate realism.
    \item We apply the \name framework to data augmentation in machine learning, showing that images with high realism score contribute to model training, and vice versa.
    \item We apply the \name framework to benchmark state-of-the-art T2I models, providing insights in their strengths and weaknesses.
\end{itemize}

%% file: latex/related_work.tex
\section{Related Work}

\subsection{Text-to-Image Generation Models}
The first generation of T2I models are based on Generative Adversarial Networks (GANs) \cite{goodfellow2014generative}. \citet{reed2016generative} first applied GAN to T2I generation, demonstrating its capability in producing plausible images. The original DALL-E \cite{dalle1}, introduced by OpenAI in 2021, leveeraged an autoregressive model trained on vast amount of data to generate diverse images. Most recently, however, diffusion models \cite{ho2020denoising} became the mainstream method for image generation, and powered state-of-the-art T2I models like Stable Diffusion 3 \cite{sd3}, DALL-E 3 \cite{dalle3}, and Kandinsky 3 \cite{kandinsky3}. Despite these advancements and the ability to generate photorealistic images, challenges remain in producing faithful and realistic outputs, particularly for objects and scenes with limited training data \cite{zhang2024text}.

\subsection{Evaluation of Text-to-Image Models}  
Various evaluation metrics for T2I models in different aspects have been proposed over the years. DALL-EVAL \cite{cho2023dall} measures the reasoning capabilities and social biases of T2I models. CommonSense-T2I \cite{fu2024commonsense} evaluates the commonsense knowledge of these models. A notable family of metrics are on output faithfulness, namely the alignment between text prompt and generated image. CLIP-Based metrics like CLIPScore \cite{hessel2021clipscore} and CLIP-R \cite{park2021benchmark} leverage the CLIP model \cite{clip} to compute embeddings for both the text and the image, then returns the cosine similarity between the embeddings. More recently, VQA-based methods such as TIFA \cite{hu2023tifa} employ visual question-answering (VQA) models to answer structured questions generated from the text prompt, arriving at a faithfulness score. On top of it, Davidsonian scene graph \cite{chodavidsonian} formalizes the properties of these structured questions to ensure consistency and reliability in evaluations. Despite much progress in faithfulness metrics, reference-free evaluation that focus on image quality, especially image realism, remains largely unexplored. Inception Score \cite{salimans2016improved} and Fréchet inception distance \cite{heusel2017gans} are two widely-used image quality metrics, but they require ground truth images, and rely on a relatively small pre-trained classifier that is not suitable for complex datasets \cite{frolov2021adversarial}. This gap highlights the need for visual realism assessments.

%% file: latex/method.tex
\section{Method}
We study image realism in three dimensions: correctness of fine-grained visual attributes, plausibility of unusual visual relationships, and realistic visual styles. \name evaluates each dimension with a separate module.

\subsection{Evaluation of Visual Attributes}
\name's first dimension of evaluation is the correctness of visual attributes. T2I models, when tasked with producing specific objects such as a particular animal species, are prone to generating outputs with missing or inaccurate attributes \cite{huang2024t2i, parihar2024precisecontrol}, which degrade output quality from a realism perspective. Inspired by the Davidsonian scene graph~\cite{chodavidsonian}, we design a graphical schema using a series of atomic and unique questions to verify the presence and accuracy of each visual attribute.

The process begins with an existence check to determine whether the target object is present in the image. The prompt \texttt{"Is there a realistic animal or plant in the image?"}, is sent to a VQA model for verifying. If the response is negative, the evaluation concludes with a score of zero for the image. Otherwise, the framework evaluates specific attributes based on a pre-defined schema $\mathcal{S}$, where each attribute $a \in \mathcal{S}$ consists of a pair $(p, d)$, representing the attribute's part name and description, respectively. The schema can be automatically generated using a LLM with access to a knowledge base, and we provide more details on the different ways we use to generate schema for each dataset in \S\ref{dataset}.

For each attribute $a_i = (p_i, d_i)$, the framework performs two checks:
1. A visibility check, where the model determines whether the part $p_i$ is discernible in the image using a prompt like \texttt{"Can you see the $p_i$?"}. Let $V_i \in \{0, 1\}$ denote the result, where $V_i = 1$ if $p_i$ is visible.
2. A description match check, performed only if $V_i = 1$, where the model verifies whether the part's appearance matches its description $d_i$ using a prompt like \texttt{"Is the $p_i$ $d_i$?"}. Let $M_i \in \{0, 1\}$ denote the result, and $M_i = 1$ if the description matches. The framework computes two key metrics: the confidence score $C$ defined as the total number of visible attributes, and the realism score $R$ defined as the total number of visible attributes that are correctly depicted. The notations are as follows, where $N$ denotes the total number of attributes in $\mathcal{S}$. \\
\[
C = \sum_{i=1}^N V_i, \quad R = \sum_{i=1}^N V_i \cdot M_i.
\]
Finally, the normalized attribute score is computed as the ratio between realism and confidence scores:
\[
S_{att} = \begin{cases} 
\frac{R}{C}, & \text{if } C > 0, \\
0, & \text{if } C = 0.
\end{cases}
\]

\subsection{Evaluation of Visual Relations}
The second dimension of evaluation focuses on the realism for different objects within an image, and the visual relations between them. T2I models often fail with generating unusual visual relationships \cite{gokhale2022benchmarking}, and even if the relationship is correctly depicted, the objects may look unreal as depicted in Figures~\ref{fig:teaser} and \ref{fig:method}. Therefore, we evaluate whether the objects and  relationships between them are both present and realistic.

Given a query that specifies relationships among objects, the framework evaluates each entity $e$ and the relationship $r$, where the relationship between entities $e_i$ and $e_j$ is expressed as $r_{ij}$. For each entity, two prompts are issued: one for visibility: \texttt{"Can you see a $e$?"}, and another for realism: \texttt{"Is the $e$ realistic and natural?"}. If any object is found to be missing, the evaluation concludes with a score of zero for the image. If all objects are present, each relationship is evaluated using a separate prompt, \texttt{"Can you see the $e_i$ $r_{ij}$ $e_j$?"}. The final relationship score is computed as:
\[
S_{rel} = \sum_{i=1}^N (V_i + M_i) + \sum_{i,j=1, i\neq j}^N{R_{ij}},
\]
where $V$, $M$, and $R$ refer to visibility, realism, and relationship checks, respectively, and $N$ is the total number of objects in the query.

\subsection{Evaluation of Visual Styles}
The last dimension of evaluation is the realism of visual styles. Even if the prompt asks for a photorealistic image, T2I models often produce outputs with inconsistent styles, especially for uncommon objects or unusual relationships. The outputs typically have an illustrative or cartoonish appearance, deviating sharply from the realism requirement, as shown in Figure~\ref{fig:teaser} and \ref{fig:method}

To assess the style realism of an image, we fine-tune the CLIP model~\cite{clip} on two classes of images: "photo" and "illustration." We create a fine-tuning dataset of 9,400 images, where the realistic ones are randomly sampled from the iNaturalist, Birds, and UnRel datasets, and the illustrative ones are evenly generated by the four T2I models using the prompt \texttt{"an illustration of \{CONTENT\}"}. The fine-tuning process is configured with a constrative learning objective, a learning rate of $5 \times 10^{-5}$, a batch size of 8, and five training epochs. During evaluation, an input image is processed through the fine-tuned model, which outputs a probability score indicating the likelihood of the image belonging to the "photo" class, which we use as the realism score.

By combining scores for attribute and relationship with style, \name~provides a comprehensive framework for evaluating the realism of T2I-generated images.

%% file: latex/experiments.tex
\section{Experiments}
We demonstrate \name's alignment with human judgment and applicability to three machine learning applications, and arrive at a realism benchmark for current T2I models. While different VQA model can be used, we choose GPT-4o for its state-of-the-art vision capability, scoring 69.1 on MMLU \cite{openai2024gpt4o}. A comparison for different VQA models is in \S\ref{sec:model}. Copyright information on scientific artifacts used are in Appendix \ref{sec:copyright}. 

\subsection{Datasets}
\label{dataset}
To evaluate \name's capability in scoring image realism, we experiment with three datasets:
\begin{itemize}
    \item iNaturalist \cite{van2018inaturalist} is a challenging dataset of natural objects. It contains 10,000 fine-grained classes of animals, plants, and fungi, and we randomly sample 200 classes for attribute and style evaluation. The dataset does not come with an attribute schema, but we create the schema for each class by crawling its "Description" column from Wikipedia, then using a GPT-4 model to extract and summarize the major parts and corresponding descriptions.
    \item Birds \cite{WahCUB_200_2011} is another fine-grained dataset containing 200 classes of bird species. Every image is annotated with 312 binary visual annotations covering features such as color and shape. We group images of the same class, and gather common annotations as attribute schema for the class. We use all 200 classes for attribute and style evaluation.
    \item UnRel \cite{peyre2017weakly} is a scene graph dataset with unusual visual relationships such as "a person carrying a bed." The number of unique object-relationship-object triplets is 76. The low resource nature of the images brings challenges for T2I models to generate high quality augmented data. We use it for visual relation and style evaluation.
\end{itemize}

\begin{table*}[ht!]
\centering
\begin{tabular}{lcccccc}
\toprule
\multirow{2}{*}{Eval Method} & \multicolumn{2}{c}{iNaturalist} & \multicolumn{2}{c}{Birds} & \multicolumn{2}{c}{UnRel} \\ 
  & Spearman's $\rho$ & Kendall's $\tau$ & Spearman's $\rho$ & Kendall's $\tau$ & Spearman's $\rho$ & Kendall's $\tau$ \\ \midrule
SPICE & 0.0846 & 0.0596 & 0.2011 & 0.1541 & 0.2239 & 0.1758 \\
CLIP Score & 0.2176 & 0.1590 & 0.1698 & 0.1167 & 0.1670 & 0.1255 \\
GPT Score & 0.2716 & 0.2175 & 0.1106 & 0.0816 & 0.2092 & 0.1817 \\
\name & \textbf{0.5223} & \textbf{0.4281} & \textbf{0.6162} & \textbf{0.4880} & \textbf{0.5672} & \textbf{0.5034} \\ \bottomrule
\end{tabular}
\caption{Alignment with human judgment for \name and the baseline metrics on iNaturalist, Birds, and UnRel datasets. \name exhibits highest correlation, with a Spearman's $\rho$ metric of up to 0.62.}
\label{tab:correlation}
\end{table*}

\subsection{Text-to-Image models}
To generate diverse and high-quality synthetic data, we utilize four T2I models from three different model families. Stable Diffusion 1.1 \cite{sd1} is an early open-source latent diffusion model from Stability AI, trained on the LAION dataset. Stable Diffusion 3.5 \cite{sd3} Turbo is a more advanced iteration featuring enhanced image quality and efficiency through a diffusion transformer. DALL-E 3 \cite{dalle3}, a proprietary model by OpenAI, leverages a diffusion-based architecture and private datasets for improved caption alignment. Kandinsky 3 \cite{kandinsky3} is an open-source latent diffusion U-Net model trained on private datasets with a focus on Russian cultural elements.


\subsection{Alignment with Ground Truth}
To demonstrate the effectiveness of \name's evaluation schema, we first evaluate its alignment with ground-truth human judgment. We randomly select 100 images from each dataset, forming a subset of 300 images, and leverage the Amazon Mechanical Turk (MTurk) platform to gather human annotations. Details on worker demographics and evaluation format can be found in Appendix \ref{sec:mturk}.

We conduct the MTurk evaluation as follows: for each image, we manually summarize related attributes or relations, and craft questions on their realism. Then, each question is presented to three human workers for "yes" or "no" labels. The majority vote for each question is taken as the ground truth label. Finally, we compute a score for each image as the ratio between the number of positive labels and all questions. We calculate Spearman’s $\rho$ and Kendall’s $\tau$ scores to measure the correlation between ground truth score and the scores generated by \name end-to-end, and compare performance against three baseline methods:

\begin{itemize}
    \item SPICE: Caption-based evaluation proposed by \citet{hong2018inferring}. For each image, a caption is generated using the BLIP-Image-Captioning-Base model, and SPICE \cite{anderson2016spice}, a popular metric for caption evaluation, calculates the correlation between generated and ground truth caption, which is: "A realistic image of a \{NAME\}."
    \item CLIPScore \cite{hessel2021clipscore}: This method uses the CLIP-ViT-Base model to compute embedding for both the image and its caption, then returns their cosine similarity. The caption is: "A realistic image of a \{NAME\}."
    \item GPT Score: This method directly prompts GPT-4o, the VQA model used in \name, with the identical fine-grained knowledge base but without adhering to \name's evaluation schema. The prompt is: "This is an image of a \{NAME\}. Assess the realism of the image based on the description: \{DESC\}. Each correctly depicted and clearly visible attribute earns 1 point. Output the total score."
\end{itemize}

As shown in Table~\ref{tab:correlation}, \name exhibits better alignment with human judgment than the baselines, demonstrating superior realism reflection. The T2I alignment metrics, SPICE and CLIPScore, does not adapt well to realism evaluation, as reflected in their relatively low alignment scores. Despite leveraging the same attribute information and the GPT-4o model, \name improves Spearman’s $\rho$ by 37.1\% and Kendall’s $\tau$ by 31.3\% over the GPT Score, highlighting the effectiveness of \name’s question schema in guiding the VQA model.

\subsection{Applicability for Machine Learning Tasks}
To demonstrate the practical applicability of \name, we apply it to score and filter augmented data for three machine learning tasks: image classification, image captioning, and visual relationship detection. To ensure deterministic results, all model weights are initialized with a fixed seed, and we set sampling temperature for the VQA model to 0.

\subsubsection{Image Classification}
We evaluate image classification using the ViT-Base-Patch16-224 \cite{vit} model on the iNaturalist and Birds datasets. For each dataset, we construct four training sets: none, low-quality, random, and high-quality augmentation. Each dataset consists of 200 classes, with five randomly sampled non-synthetic images per class forming both the testing and "no augmentation" training sets. From the remaining images (synthetic and non-synthetic), we compute attribute ($S_{att}$) and style ($S_{sty}$) scores using \name, combining them via the heuristic $S_{att} \times S_{sty}$. The contribution of the style score is analyzed in an ablation study in \S\ref{ablation1}. Images are ranked by their combined score, with the top five per class forming the "high-quality augmentation" set, the lowest five forming the "low-quality augmentation" set, and five randomly selected images forming the "random augmentation" set. 

We train the classification model on each set separately for 10 epochs (20 epochs for the "no augmentation" set, as it has half the number of images per class), using a batch size of \texttt{8}, an initial learning rate of \texttt{5e-5} and a weight decay of \texttt{0.01} on a single GPU with 24GB memory.

\begin{figure}
    \begin{minipage}{0.45\columnwidth}
        \centering
        \includegraphics[width=\linewidth]{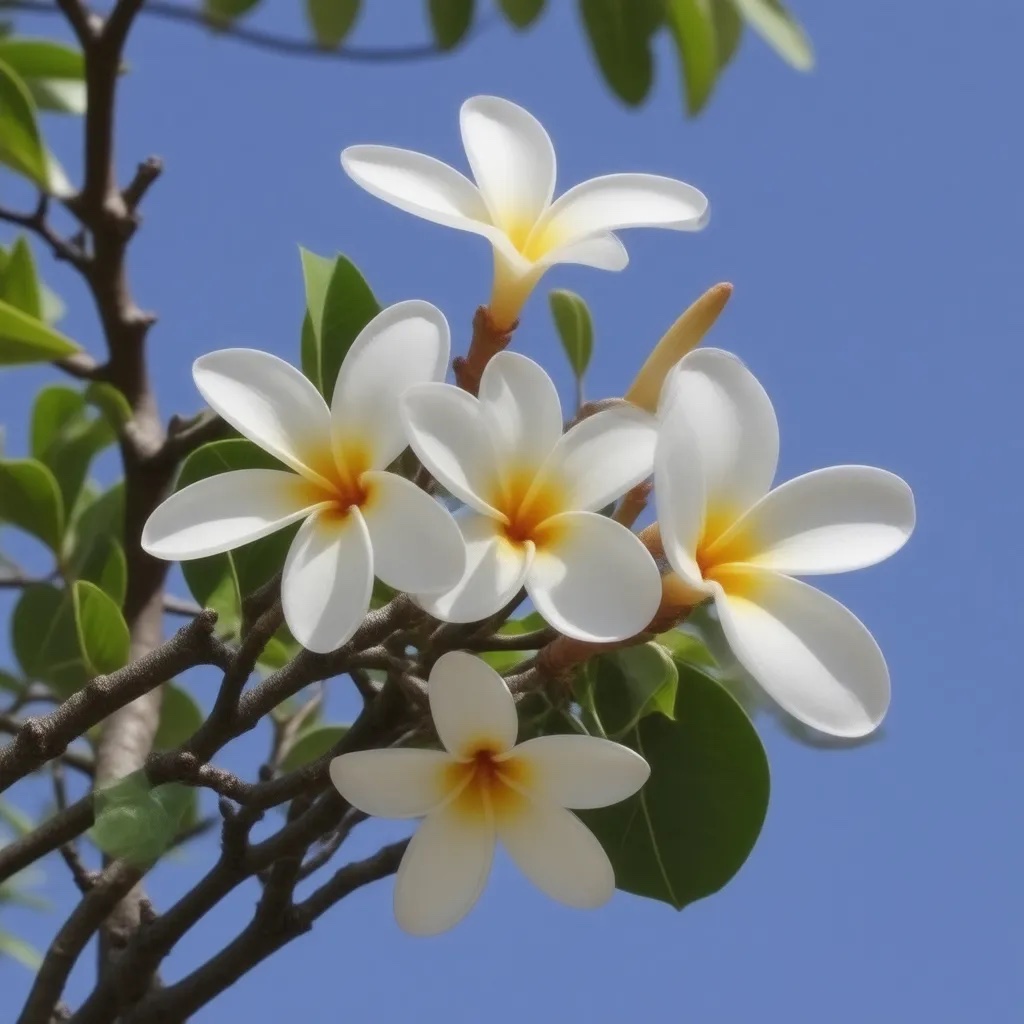}
    \end{minipage}
    \hfill
    \begin{minipage}{0.45\columnwidth}
        \centering
        \includegraphics[width=\linewidth]{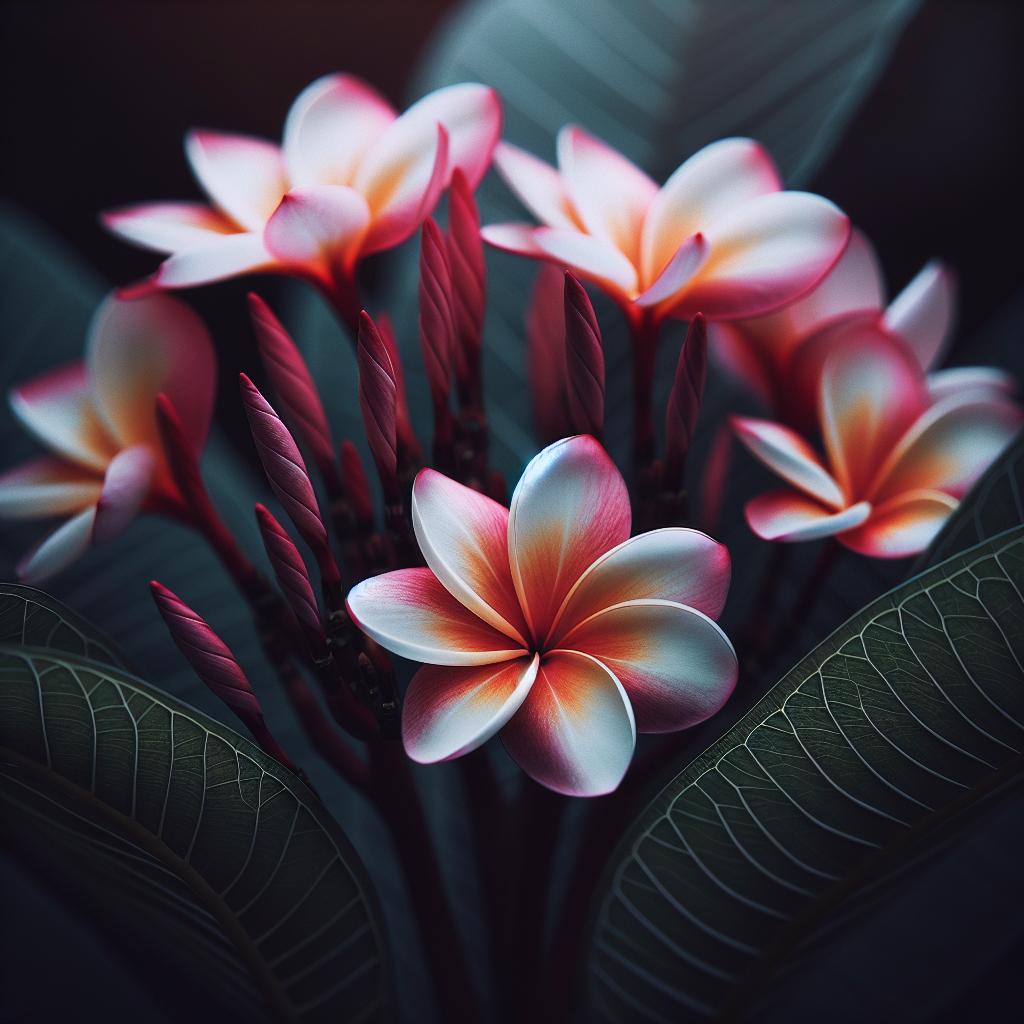}
    \end{minipage}
    \caption{Examples of high and low quality images.}
    \label{fig:example}
\end{figure}

Models trained with high-scoring images outperform the low-quality and random augmented ones by 10.6\% and 4.28\% in F1 score on average, as shown in Table~\ref{tab:classification}. In addition, we plot test performance against training steps in Figure~\ref{fig:inaturalist_f1}, providing a temporal view of the results. Importantly, we notice that models with low-scoring images may perform even worse than the ones without. These images either exhibit incorrect visual traits that mislead the classifier, or shows unrealistic styles, which may be used as an unreliable feature for real-world image classification. Figure~\ref{fig:example} shows a pair of examples, with a high-quality synthetic image on the left and a low-quality one on the right, whose visual attributes (color, petal number, leaf size) and style both deviate from the realistic requirement. \name's scoring effectively separates these.


\begin{table}[h!]
\centering
\begin{tabular}{llcc}
\toprule
Dataset & Aug Method & Accuracy & F1 \\ \midrule
\multirow{4}{*}{iNaturalist} & none & 0.6950 & 0.6937 \\
 & low quality & 0.6450 & 0.6442 \\
 & random & 0.7200 & 0.7271 \\
 & high quality & \textbf{0.8100} & \textbf{0.8070} \\ \midrule
\multirow{4}{*}{Birds} & none & 0.6700 & 0.6649 \\
 & low quality & 0.6750 & 0.6685 \\
 & random & 0.7100 & 0.7112 \\
 & high quality & \textbf{0.7200} & \textbf{0.7170} \\ \bottomrule
\end{tabular}
\caption{Image classification training results on high-quality, random, low-quality, and no augmentation training sets according to \name scores. The model trained on high-quality sets exhibits the highest accuracy.}
\label{tab:classification}
\end{table}

\subsubsection{Image Captioning}
We evaluate image captioning using the BLIP-Image-Captioning-Base \cite{blip} model on the iNaturalist and Birds datasets. The four training sets are constructed similarly to classification, except each image is assigned a caption in the format \texttt{"a photo of a \{CLASS\_NAME\}"}. The model is trained for 10 epochs with a batch size of \texttt{8}, an initial learning rate of \texttt{1e-5}, and a weight decay of \texttt{0.01} on a single GPU with 24GB memory.

\begin{figure*}[ht!]
    \centering
    \begin{minipage}{0.32\textwidth}
        \centering
        \includegraphics[width=\textwidth]{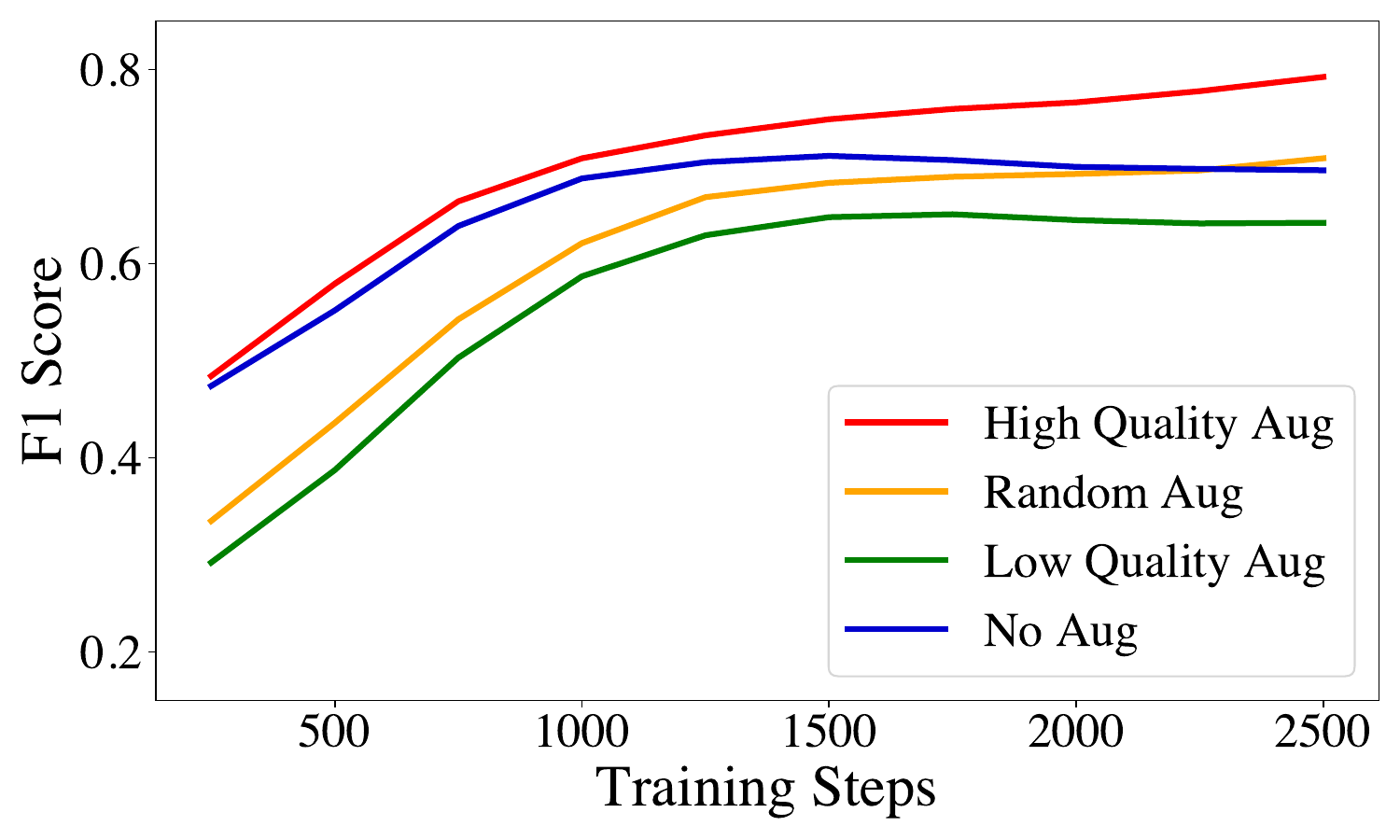}
        \caption{image classification on iNaturalist dataset (F1 score)}
        \label{fig:inaturalist_f1}
    \end{minipage}
    \hfill
    \begin{minipage}{0.32\textwidth}
        \centering
        \includegraphics[width=\textwidth]{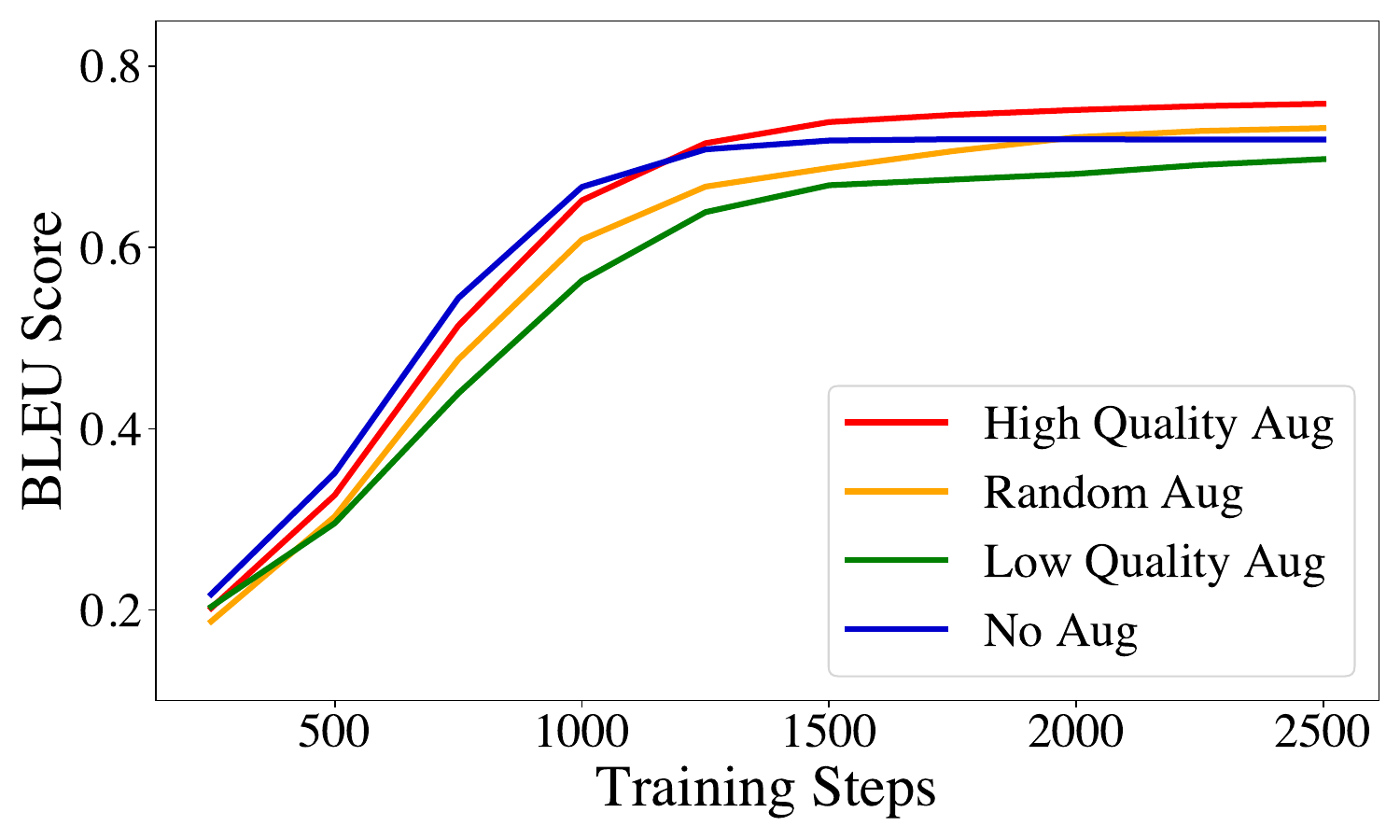}
        \caption{image captioning on iNaturalist dataset (BLEU score)}
        \label{fig:birds_bleu}
    \end{minipage}
    \hfill
    \centering
    \begin{minipage}{0.32\textwidth}
        \centering
        \includegraphics[width=\textwidth]{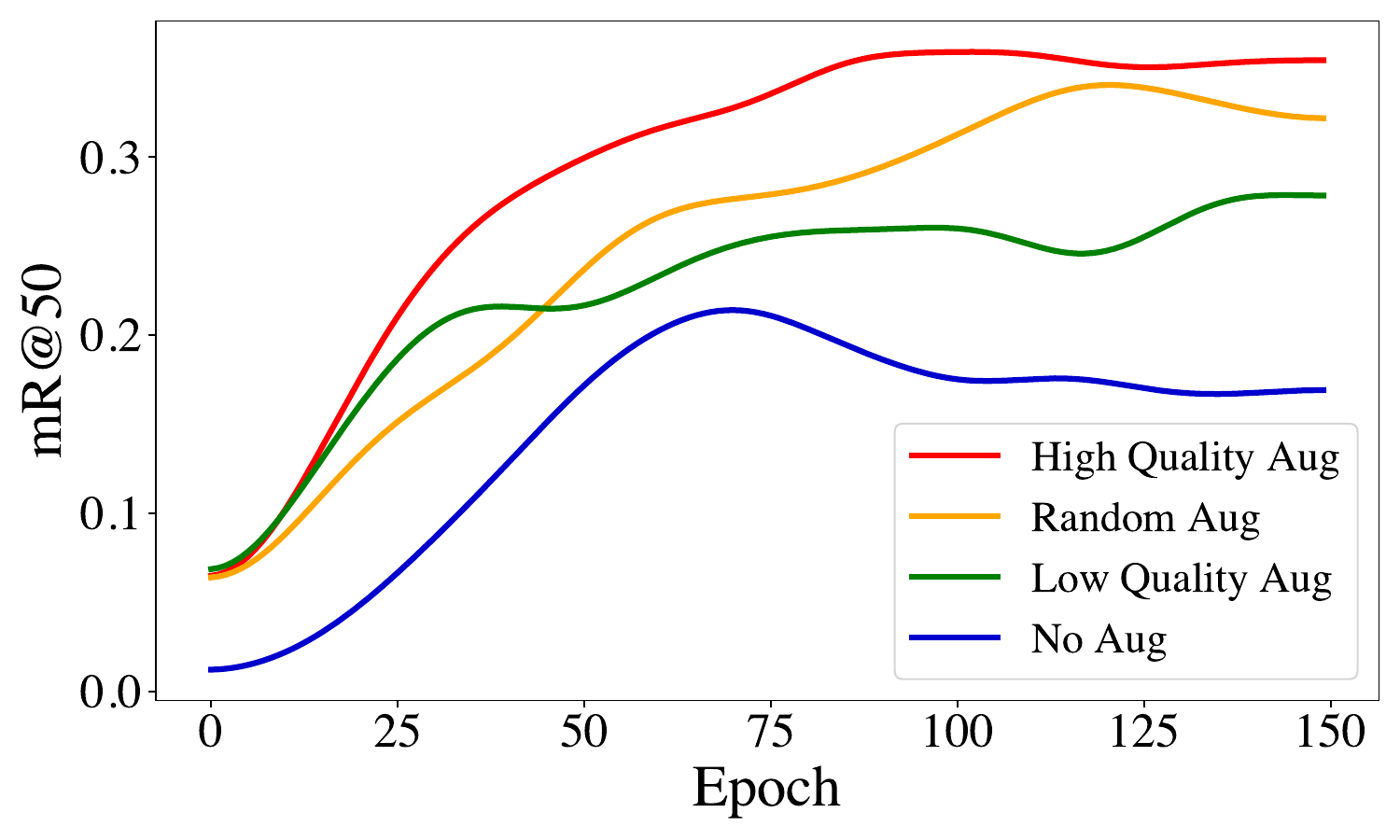}
        \caption{visual relationship detection on UnRel dataset (mR@50)}
        \label{fig:mR50}
    \end{minipage}
\end{figure*}

We show training results in Table~\ref{tab:captioning} measured by ROUGE 1 \cite{lin2004rouge} and BLEU \cite{papineni2002bleu}. ROUGE 1 measure the unigram overlap between generated and reference captions, and BLEU computes n-gram precision with a brevity penalty that penalizes candidates that are too short relative to the reference. The high-quality augmented model outperforms the random and low-quality ones with an average of 3.88\% and 5.63\% in BLEU respectively. Despite the high-quality model did not converge the fastest, as shown in the temporal plot in Figure~\ref{fig:birds_bleu}, it achieves the highest accuracy. This is because \name-selected high-quality images provide rich real-world features that takes time to learn, but eventually resulting in the model's robustness to unseen data.


\begin{table}[h!]
\centering
\begin{tabular}{@{}lccc@{}}
\toprule
Dataset & Aug Method & ROUGE 1 & BLEU \\ \midrule
\multirow{4}{*}{iNaturalist} & none & 0.8890 & 0.7194 \\
 & low quality & 0.8859 & 0.7008 \\
 & random & 0.8942 & 0.7345 \\
 & high quality & \textbf{0.9084} & \textbf{0.7612} \\ \midrule
\multirow{4}{*}{Birds} & none & 0.8753 & 0.6454 \\
 & low quality & 0.8851 & 0.6624 \\
 & random & 0.8866 & 0.6635 \\
 & high quality & \textbf{0.9042} & \textbf{0.7143} \\ \bottomrule
\end{tabular}
\caption{Image captioning training results on high-quality, low-quality, and no augmentation training sets according to \name scores. The model trained on high-quality sets exhibits the highest accuracy.}
\label{tab:captioning}
\end{table}

\subsubsection{Visual Relationship Detection}
We evaluate visual relationship detection using the Relation Transformer for Scene Graph Generation (RelTR) model \cite{reltr}. RelTR combines an object detection model for entity localization with a relational transformer that generates a sparse scene graph in a single step, predicting relationships in an object–relationship–object triplet. 

We use \name's relationship and style scores to construct four training sets from the UnRel dataset, then fine-tune and compare the RelTR model, pre-trained on the Visual Genome dataset by the original authors, across these sets. As the UnRel dataset lacks object bounding boxes, we generate them using the YOLO 11 model \cite{yolo11}. Following the authors' setup, we train RelTR for 150 epochs with a batch size of \texttt{2} and an initial learning rate of \texttt{1e-4} on four GPUs with 24GB memory each.

We show training results in Table~\ref{tab:vrd} and plot test performance in Figure~\ref{fig:mR50}. We measure relationship detection performance by mean recall at 20 and 50, which calculates the proportion of relevant relationships among top 20 and 50 recommendations, respectively. The model trained on high-quality training set achieves 7.42\% higher mR@20 and 5.32\% higher mR@50 than the low-quality one, which again highlights \name's applicability to a wide range of machine learning tasks.


\begin{table}[h!]
\centering
\begin{tabular}{@{}lccc@{}}
\toprule
Dataset & Aug Method & mR@20 & mR@50 \\ \midrule
\multirow{3}{*}{UnRel} & none & 0.1036 & 0.1821 \\
 & low quality & 0.2199 & 0.2787 \\
 & Random & 0.3123 & 0.3137 \\
 & high quality & \textbf{0.3529} & \textbf{0.3613} \\ \bottomrule
\end{tabular}
\caption{Visual relationship detection training results on high-quality, low-quality, and no augmentation training sets according to \name scores. The model trained on high-quality sets exhibits the highest accuracy.}
\label{tab:vrd}
\end{table}

\subsection{Realism Benchmark for T2I Models}
As \name is effective in rating augmented data, we propose using its output score to benchmark current T2I models, and show results in Table~\ref{tab:benchmark}. The benchmark composes of four scores: attribute, relationship, style, and average. While DALL-E 3 has the highest relationship score among the four T2I models, it is prone to generating illustration-style images and has the lowest style score. Stable Diffusion v3.5 exhibits higher attribute and relationship scores than its predecessor, yet its style score is lower than it. Kandinsky 3 generates images with the highest degree of style realism but its knowledge of fine-grained object attributes is relatively lacking. On average, however, \name rates Kandinsky 3 to be the best-performing among the four models in terms of output realism. We believe that the benchmark provides novel yet overlooked aspects of evaluation for current T2I models, and the results would benefit the whole community.

\begin{table*}[ht!]
\centering
\begin{tabular}{lcccc}
\toprule
T2I Model & Attribute Score & Relationship Score & Style Score & Average Score \\ \midrule
DALL-E 3 & 0.5475 & \textbf{0.7827} & 0.2430 & 0.5244 \\
SD v1.1 & 0.5717 & 0.3739 & 0.6356 & 0.5271 \\
SD v3.5 & \textbf{0.5791} & 0.7315 & 0.5380 & 0.6162 \\
Kandinsky 3 & 0.4925 & 0.7301 & \textbf{0.6745} & \textbf{0.6324} \\ \bottomrule
\end{tabular}
\caption{Realism benchmark for four popular T2I models in the three dimensions. Kandinsky 3 achieves the highest average score while DALL-E 3 has the lowest due to unrealistic output styles.}
\label{tab:benchmark}
\end{table*}

%% file: latex/discussion.tex
\section{Discussion}

\subsection{Performance of Different VQA Models}
\label{sec:model}
We conduct an ablation study to study the effect of using different VQA models for \name evaluation. While GPT-4o is our default VQA model, we compare the performance of four other models: BLIP2-Flan-T5-XL \cite{li2023blip}, PaliGemma-3B \cite{steiner2024paligemma}, mPLUG-Owl3-7B \cite{ye2024mplug}, and Gemini-1.5-Flash-002 \cite{team2024gemini}. The sampling temperature is always set to 0. We evaluate the models by alignment with human judgment on iNaturalist dataset, and show results in Table~\ref{tab:ablation_vqa}. We observe that the performance of commercial models (Gemini and GPT) are stronger than the open-source ones, and GPT-4o has the highest correlation with human labels. mPLUG is the recommended open-source model as it offers a correlation value close to Gemini, with a difference of 1.27\% in Kendall's $\tau$. BLIP2 exhibits poor performance mainly because it is prone to false positives, namely answering "yes" when the ground truth answer is "no". Specifically, its average score for all test images is 0.63, whereas that provided by GPT is 0.57. With the rapid improvements of T2I models, we expect \name's accuracy to further improve in the future.

\begin{table}[ht!]
\centering
\resizebox{0.45\textwidth}{!}{
\begin{tabular}{lccc}
\toprule
VQA Model & Spearman's $\rho$ & Kendall's $\tau$ \\ \midrule
BLIP2 & 0.0255 & 0.0132 \\
PaLI & 0.3336 & 0.2845 \\
mPLUG & 0.4733 & 0.3969 \\
Gemini & 0.4950 & 0.4096 \\
GPT-4o & \textbf{0.5223} & \textbf{0.4281} \\ \bottomrule
\end{tabular}
}
\caption{Comparison of \name performance using different VQA models, measured by alignment with human judgment on the iNaturalist dataset.}
\label{tab:ablation_vqa}
\end{table}

\subsection{Contribution of Style Score in Augmented Data Ranking}
\label{ablation1}
To demonstrate the contribution of incorporating the style score in addition to attribute and relationship scores for ranking augmented images, we conduct an ablation study on the iNaturalist dataset. The results in Table~\ref{tab:ablation_style} show that when training the image classification model, the high-quality training set achieves a 3.70\% higher F1 score when ranked using the combined attribute and style scores, compared to using the attribute score alone. This indicates that the combined score effectively filters images with realistic styles, leading to improved model performance.

\begin{table}[ht!]
\centering
\begin{tabular}{@{}lcc@{}}
\toprule
Aug Setting & Accuracy & F1  \\ \midrule
Attribute Only & 0.7700 & 0.7700 \\
Attribute + Style & \textbf{0.8100} & \textbf{0.8070} \\ \bottomrule
\end{tabular}
\caption{Ablation study on \name's use of style score in addition to attribute score for image classification.}
\label{tab:ablation_style}
\end{table}

\subsection{Effect of Fine-Tuning in Visual Style Evaluation}
In order to show that our fine-tuned CLIP models better evaluates style realism than the out-of-the-box one, we conduct an ablation study using 100 real and illustrated-styled images on iNaturalist dataset, and show results in Table~\ref{tab:ablation_style}. After fine-tuning, the Spearman's $\rho$ correlation with ground truth labels increases by 4.92\% and the Kendall's $\tau$ metric increases by 4.05\%, demonstrating the effectiveness of our style dataset.

\begin{table}[ht!]
\centering
\resizebox{\columnwidth}{!}{
\begin{tabular}{@{}lccc@{}}
\toprule
Setting & Spearman's $\rho$ & Kendall's $\tau$ \\ \midrule
W/o Fine-Tuning & 0.7775 & 0.6349 \\
w/ Fine-Tuning & \textbf{0.8267} & \textbf{0.6754} \\ \bottomrule
\end{tabular}
}
\caption{Comparison of style evaluation results with and without fine-tuning the CLIP model, measured by alignment with human judgment on the iNaturalist dadtaset.}
\label{tab:ablation_ft}
\end{table}

%% file: latex/conclusion.tex
\section{Conclusion}
We introduce \name, a framework to evaluate the realism of T2I model outputs across attributes, relationships, and styles. \name aligns closely with human judgment and improves machine learning tasks like classification, captioning, and relationship detection by filtering high-quality data. Our benchmarks offer insights into current T2I models and highlight directions for future advancements.

%% file: latex/limitations.tex
\section{Limitations}

\name requires a LLM for summarizing attribute information from the knowledge base, and a VQA model for question answering. Using commercial models can lead to financial cost, as it takes \$0.03 to evaluate an image on average with GPT-4o. However, this can be eliminated using open source models such as mPLUG-Owl3, which already show similar performance to Google's Gemini-1.5 model. Also, it takes 10 seconds on average to score an image, but processing speed can be greatly accelerated using asynchronous programming. In the future we should cover more dimensions, such as logic, commonsense knowledge, etc.


%% file: latex/appendix.tex
\section{Appendix}

\subsection{Scientific Artifacts}
\label{sec:copyright}
We list the name, source, and license for each of the scientific artifact we use below.

\begin{table}[h]
    \centering
    \resizebox{\columnwidth}{!}{
    \begin{tabular}{l l l}
        \hline
        \textbf{Name} & \textbf{Source} & \textbf{License} \\
        \hline
        iNaturalist & \citet{van2018inaturalist} & iNaturalist terms of service \\
        Birds & \citet{WahCUB_200_2011} & Open to non-commercial research \\
        UnRel & \citet{peyre2017weakly} & UnRel code and data license \\
        SPICE & \citet{hong2018inferring} & GNU Affero General license v3.0 \\
        CLIPScore & \citet{hessel2021clipscore} & MIT license \\
        ROUGE & \citet{lin2004rouge} & Apache License v2.0 \\
        BLEU & \citet{papineni2002bleu} & Apache License v2.0 \\
        \hline
    \end{tabular}
    }
    \caption{Scientific artifact sources and licenses.}
    \label{tab:dataset_licenses}
\end{table}

\subsection{MTurk Details}
\label{sec:mturk}
\begin{figure}[h!]
    \centering
    \includegraphics[width=\columnwidth]{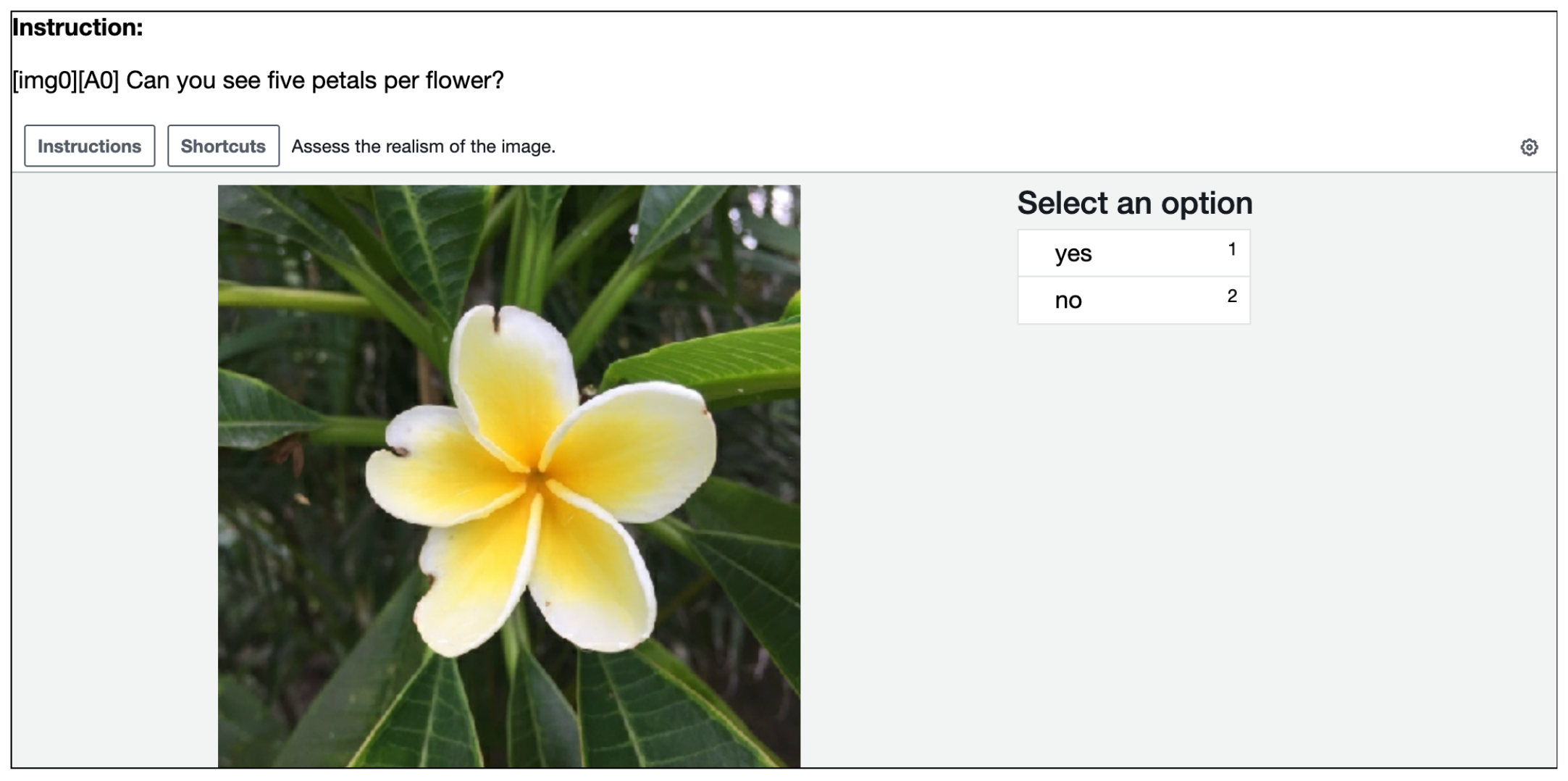}
    \caption{An example question that is sent to workers.}
    \label{fig:MTurk}
\end{figure}

The workers are from diverse English-speaking backgrounds, and all of them are MTurk Master qualifiers. They were paid \$0.05 for each example. based on the average completion time for each task, the estimated wage rate is \$9/hour, which is higher than the US minimum wage (\$7.25/hour). All data we use are granted for research purposes. Figure~\ref{fig:MTurk} shows the interface for our MTurk evaluation.

